\documentclass[12pt]{iopart}

\usepackage{graphicx}
\expandafter\let\csname equation*\endcsname\relax
\expandafter\let\csname endequation*\endcsname\relax
\usepackage{amsmath}
\usepackage{url}
\usepackage[dvipsnames]{xcolor}
\begin{document}
\bibliographystyle{vancouver}

\title{Machine Learning for Magnetic Phase Diagrams and Inverse Scattering Problems}

\author{Anjana M. Samarakoon$^1$\footnote{Notice: This manuscript has been authored by UT-Battelle, LLC, under contract DE-AC05-00OR22725 with the US Department of Energy (DOE). The US government retains and the publisher, by accepting the article for publication, acknowledges that the US government retains a nonexclusive, paid-up, irrevocable, worldwide license to publish or reproduce the published form of this manuscript, or allow others to do so, for US government purposes. DOE will provide public access to these results of federally sponsored research in accordance with the DOE Public Access Plan (http://energy.gov/downloads/doe-public-access-plan).} 
and D. Alan Tennant$^{2,3,4}$}

\address{$^1$ Neutron Scattering Division, Oak Ridge National Laboratory, Oak Ridge, TN 37831 USA}
\address{$^2$ Materials Science and Technology Division,}
\address{$^3$ Shull Wollan Center - A Joint Institute for Neutron Sciences,\\
Oak Ridge National Laboratory, Oak Ridge, TN 37831 USA}
\address{$^4$ Quantum Science Center, Oak Ridge TN 37831 USA}
\ead{samarakoonam@ornl.gov, tennantda@ornl.gov}

\begin{abstract}
Machine learning promises to deliver powerful new approaches to neutron scattering from magnetic materials. Large scale simulations provide the means to realise this with approaches including spin-wave, Landau Lifshitz, and Monte Carlo methods. These approaches are shown to be effective at simulating magnetic structures and dynamics in a wide range of materials. Using large numbers of simulations the effectiveness of machine learning approaches are assessed. Principal component analysis and nonlinear autoencoders are considered with the latter found to provide a high degree of compression and to be highly suited to neutron scattering problems. Agglomerative heirarchical clustering in the latent space is shown to be effective at extracting phase diagrams of behavior and features in an automated way that aid understanding and interpretation. The autoencoders are also well suited to optimizing model parameters and were found to be highly advantageous over conventional fitting approaches including being tolerant of artifacts in untreated data. The potential of machine learning to automate complex data analysis tasks including the inversion of neutron scattering data into models and the processing of large volumes of multidimensional data is assessed. Directions for future developments are considered and machine learning argued to have high potential for impact on neutron science generally. 
\end{abstract}

%
\vspace{2pc}

%
\submitto{\JPCM (special edition on Machine Learning, ed. Jorge Quintanilla et al.)} 
%
%
%

\section{Introduction}

Machine learning (ML) combined with high-performance simulations provides new opportunities for understanding complex behaviour and large experimental data sets \cite{stfc_ml,osti_1579323}. Magnetism and its exploration using neutron scattering is a notable target for such applications because of the large data volumes and simulations involved. Further, there is increasing interest in magnetic materials for information and quantum technologies such as candidate spin liquid materials \cite{spin_liquids} but these are proving to be difficult to analyze with current methods. 
Specifically, magnetic interactions in the solid state come in a wide range of couplings, both short- and long-range, that are challenging to identify and extract from experimental data. Approaches are needed that automate modeling and interpretation of neutron scattering data, and can deal with complex phases and dynamics.

State-of-the-art neutron instrumentation collects detailed information in the form of 3- or 4-dimensional (3D reciprocal space and energy transfer) scattering data sets \cite{diffraction_suite,horace}. However, the inverse scattering problem and extraction of a microscopic model and parameters are usually ill-posed. Data analysis and modeling takes a great deal of time and expertise to accomplish; as such it is not unusual for an experienced group of researchers to take upwards of a year to understand and model one experiment. An additional difficulty concerns the uniqueness of any data analysis solution and its range of uncertainty. Any proposed model should agree with observations using different physical probes and under different parametric conditions. Co-analysis with these is then needed.

Unsupervised ML has recently been used to demonstrate the inversion of magnetic diffuse scattering data to model parameters \cite{Samarakoon:2020aa}; historically a fundamental limitation of scattering techniques \cite{ml_review}. This demonstration involved extraction of phases and dynamics from simulations as well as analyzing neutron data in an automated way. If realistic simulations are performed over a wide enough range of couplings that they encompass the potential ranges for a material then ML training should be comprehensive in covering experimental outcomes. Neural networks trained on the simulations were shown to locate model parameters from data and so inversion from the data to model was achieved. Such an approach does not depend on finding an exact solution and an effective strategy is to start with a more approximate microscopic model and map out the parameter space. Extending the model from the best region is relatively less computationally expensive compared to full exploration. An additional benefit to applying ML is that dimensionality reduction and clustering can also identify and categorizes phases in the model and experiment generating new knowledge and aiding the broader interpretation of the data.  


In this paper we consider the practical steps involved in applying machine learning to neutron scattering measurements of magnetic materials. We do this by reviewing previous work and putting it in context as well as critiquing the methodologies. The potential of these developments is assessed and example applications given.

The paper is organized as follows: In the next section the principles of large scale simulations of magnetic structures and dynamics using Monte Carlo and Landau Lifshitz approaches are explained and examples given for two magnetic systems, the two dimensional Heisenberg antiferromagnet and honeycomb Kitaev model. The following section details the application of machine learning. This covers dimensionality reduction techniques to compress large volumes of simulation data using principal component analysis (PCA) and non-linear autoencoders (NLAEs). Also addressed is the categorization of phases and generation of phase diagrams. Parameter optimization from neutron data and mapping of potential regions of fit as well as the use of generative models and their potential designs are also explained. Finally, a brief discussion of future directions is followed by a conclusion and summary, 

\section{Large scale simulations of magnetic structures and dynamics}

In magnetic insulators, spins are localized on regular arrays of sites on crystalline lattices \cite{blundell}. These have a wide range of symmetries and anisotropies that account for the remarkable array of magnetic systems and phenomena known. While specialized approaches are needed for magnetic systems with large zero-point quantum fluctuations \cite{jpsj_tennant,qmag} it is believed that many magnetic materials can be simulated with (semi-)classical methods such as spin-wave theory \cite{spin-waves,Toth_2015}; at least accurately enough to be usefully utilized with ML. Monte Carlo and Landau Lifshitz dynamics provide attractive complementary approaches to spinwaves for the classical modeling of magnetic materials as general purpose methods \cite{Huberman_2008}. They treat non-linearity in the spin configurations and dynamics as well as thermal effects and are available in some molecular dynamics packages such as LAMMPS \cite{lammps}, although here we use our own codes optimized to neutron studies of magnetism.

\subsection{Computation of spin configurations and dynamics}\label{computations}

The essential approach to calculating the spin configurations and dynamics is presented here with details given in \cite{Huberman_2008}. In the case of localized moments the individual vector spins ${\mathbf S}_i=\left[ S_i^x,S_i^y,S_i^z \right]$ are at positions $\mathbf{R}_i$ which are typically on a lattice although positional and site disorder are easily treated. The energy due to interactions is given by the spin Hamiltonian, $\mathcal{H}=\mathcal{H}(\mathcal{J},\mathbf{S},\mathbf{R})$, where $\mathcal{J}$ is the set of interaction parameters. The Hamiltonian is the master equation for the system and the time development of the system $\mathbf{S}(t)$ (Landau-Lifshitz dynamics) from any given starting configuration $\mathbf{S}(t_0)$ can be evaluated using the equations of motion which are readily derived through Poisson brackets involving the spin components and $\mathcal{H}$. Given its importance a main goal of the neutron scattering measurements is to determine $\mathcal{H}$ {\it i.e.} identify the model and its couplings $\mathcal{J}$, as well as understanding its physical behavior as verified by experiment.  

Realistic spin configurations can be prepared through annealing based on the Metropolis algorithm, a Markov Chain Monte Carlo method \cite{metropolis1953equation}. The Metropolis algorithm anneals the configuration $\mathbf{S}$ to be representative of the system in thermal equilibrium at chosen temperature $T$. From these configurations and the time development $\mathbf{S}(t)$ a full range of physical properties can be calculated. Generally multiple configurations are required to gain enough statistics to effectively evaluate the physical properties. These can be sourced by independent annealing processes or by evolving a configuration in time sufficiently that the disordering interactions in the system have created a new state uncorrelated with the first and using this as a new $t_0$.

 Two important properties to simulate are susceptibility and heat capacity. The susceptibility $\chi_m$ is calculated using the magnetization from Monte Carlo simulation with and without field; using the spin fluctuations from the dynamics simulations is another option. Meanwhile the heat capacity can be calculated from the Monte Carlo simulations using the total energy change with temperature, or alternatively its variance. As the Landau-Lifshitz dynamics involves conservation of energy of the total system (micro-canonical ensemble) either the fluctuations are calculated from Metropolis annealing at a constant temperature or by using the energy fluctuations of a sub system and treating the rest of the system as a heat bath. Note, however, that heat capacity is particularly susceptible to quantum effects and semi-classical approximations require to be used with care.

Neutron scattering is directly able to probe the dynamical correlations in the material. The scattering from the magnetic system is proportional to the cross section \cite{neutron_scattering}:
\begin{equation}
\frac{d^2\sigma}{d\hbar\omega\Omega}=\frac{k_f}{k_i}r_m^2 \sum_{\alpha,\beta} \frac{g_{\alpha}g_{\beta}}{4} \left(\delta_{\alpha\beta}-
\frac{q_{\alpha}q_{\beta}}{q^2}\right)
|F(\textbf{Q})|^2
\mathcal{S}^{\alpha\beta}\left( \textbf{Q},\omega \right)
\label{eq:neutron_scattering}
\end{equation}
where \textbf{Q} and $\hbar \omega$ are the wavevector and energy transfer in the scattering process, $k_i$ and $k_f$ are the initial and final wavevectors of the neutrons, $r_m$ is a scattering factor, $\alpha,\beta=x,y,z$ are cartesian coordinates indicating initial and final spin polarization of the neutron, $F(Q)$ is the magnetic form factor and $\mathcal{S}^{\alpha\beta}\left( \textbf{Q},\omega \right)$ is the spin correlation function. The pre-factors are set by measurement conditions or atomic properties of the material and so are straightforwardly evaluated allowing quantitative experimental determination of $\mathcal{S}\left( \textbf{Q}, \omega \right)$.  

The dynamical correlation function $\mathcal{S}^{\alpha\beta}(\textbf{Q},\omega)$ is equal to the Fourier transform of the spin-spin correlation functions in space and time and is equal to: 
\begin{equation}
\mathcal{S}^{\alpha\beta}(\textbf{Q},\omega)=\frac{1}{2\pi N}
\sum_{i,j} e^{i\textbf{Q}.\left( \textbf{R}_j-\textbf{R}_i \right)}
\int_{-\infty}^{\infty} e^{-i\omega t} \langle S_i^{\alpha}(t_0)S_j^{\beta}(t_0+t) \rangle dt.
\label{eq:correlations}
\end{equation}
For the classical ($\mathcal{C}$) simulations of spins used here the time development can be approximately calculated on an appropriately spaced 
set of discrete times from $t_0$ to $t_N$ with spacing $\Delta t$ \cite{Huberman_2008}. Averaging over multiple simulations is typically required to yield good statistics. Finally, scattering is a quantum mechanical process and the expression needs to be modified to account for detailed balance \cite{Huberman_2008} and modifications of the density of states which can be addressed as an energy dependent prefactor \cite{spinwaves_disorder}. We present simulations for two examples of dynamics in Subsection \ref{examples} below.

Energy integrated correlations are measured in diffraction experiments and appear as Bragg and diffuse scattering. The experimental data on Dy$_2$Ti$_2$O$_7$ shown in Section \ref{ml_application}, taken as an example for the application of machine learning, is from energy integrated diffraction taken at the CORELLI instrument at the Spallation Neutron Source, Oak Ridge National Laboratory \cite{Samarakoon:2020aa}. The scattering factor $S(\textbf{Q})$ is calculated from:
\begin{equation}
    \mathcal{S}(\textbf{Q})=\frac{1}{2\pi N}\left| S_Q^{\alpha} S_{-Q}^{\beta} \right|
\end{equation}
with 
\begin{equation}
    S_Q^{\alpha}=
 \sum_i S_i^{\alpha} (t_n) e^{i\textbf{Q}\cdot \textbf{R}_i}.
\end{equation}
It is noteworthy that phase information in the underlying structure and dynamics is lost in the scattering process which means that 
theories are needed to calculate and interpret the scattering from materials. 

The LL and Monte Carlo methods are well suited to big data techniques as they can provide large numbers of detailed simulations from which new insight can be extracted. In addition signal processing can be undertaken and the structures and dynamics readily visualized. An example of this is the movie in ref. \cite{Samarakoon_2017} (see Supplementary Movie 1.). A low temperature spin configuration in a highly frustrated Kitaev magnet (see example \ref{kitaev} below) is driven by flipping a single spin. The time evolution of the spin dynamics reveals two different frequency behaviors (indicated by color) which have been separated by filtering. The dynamics shows that the dynamics is confined to closed loop configurations of spins. This is related to conservation laws and topological properties of the Kitaev model on a honeycomb lattice but shows how such simulations can be used to rapidly uncover non-trivial ground state configurations and dynamics in models.

\subsection{Examples of modeling of magnetic materials}\label{examples}

To illustrate the application of LL calculations we discuss examples of simulations of neutron scattering data. 

\subsubsection{2D Heisenberg Antiferromagnet}
 
\begin{figure}
\centering\includegraphics[width=0.9\textwidth]{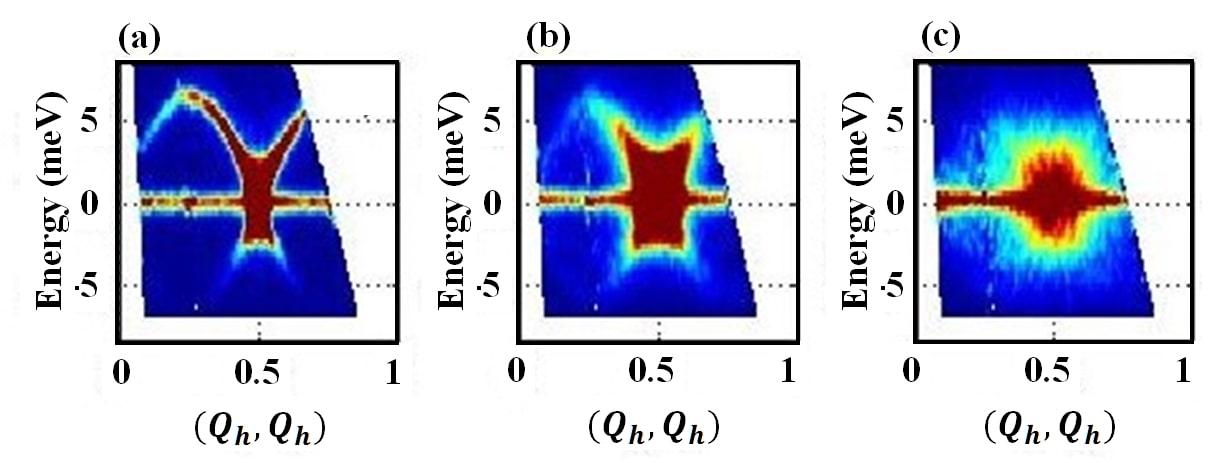}
	\caption{\textbf{Thermal development of spin waves in Rb$_2$MnF$_4$:} Cross sectional slices through the 3D volumes in $(Q_h,Q_k, \hbar \omega)$ space rendered from data collected at the MAPS spectrometer, ISIS Facility UK \cite{Huberman_2008}. Colour shading is the intensity in neutron scattering with red stronger and blue weaker. Slices are shown for temperatures a) 21.3 K which is well below the Neel ordering temperaure $T_N$, in the three-dimensionally ordered phase. Well defined transverse spin waves are observed across the Brillouin zone. b) 46.9 K, above $T_N$ but with significant short range order. Spin waves are observed although with increased lifetime broadening. The spin waves are over-damped for wave lengths longer than the correlation length. c) 100.7 K, above the Curie Weiss temperature. Over-damped behaviour alone is seen in the paramagnetic phase.  Note that these slices are taken from the measured data. No background has been subtracted and in particular incoherent scattering is seen at zero energy transfer. This figure is reproduced from \cite{Huberman_2008}.
	}
\label{flo:sqw_rb2mnf4}
\end{figure}

Comprehensive studies of magnetic excitations in the 2D spin-5/2 Heisenberg antiferromagnet Rb$_2$MnF$_4$ have been made \cite{Huberman_2005,Huberman_2008} along with an in-depth comparison with LL simulations \cite{Huberman_2008} and spin-wave theory \cite{Huberman_2005}. This covers temperature from deep in the ordered phase up to paramagnetic behavior with highly thermally disordered spins, $0.13 < k_B T/4JS < 1.4$. Well defined spin-waves are observed up to near the Curie-Weiss temperature, $\Theta_{CW}$, Figure (\ref{flo:sqw_rb2mnf4}). At lowest temperatures the accuracy of spin-wave theory is confirmed including the two-magnon cross-section which is very small \cite{Huberman_2005}. The LL simulations were found to agree with the spinwaves at lowest temperature but also provide a good description of the intermediate- and high-temperature regimes over all wave-vector and energy scales using the appropriate spin length with a crossover from quantum spin length $S$ at low temperatures to fully classical dynamics with $\sqrt{S(S+1)}$ observed around $\Theta_{CW}/S$ where the spin $S=5/2$.
Such Monte Carlo and LL calculations are able to model the phases and phase transitions too and so can provide comprehensive modeling.

\subsubsection{Honeycomb lattice}\label{kitaev}

Honeycomb lattices with anisotropic exchange interactions provide very rich systems displaying highly non-linear magnetic phenomena. We have studied the spin-$S$ Kitaev model
using Monte Carlo simulations combined with LL semi-classical spin dynamics \cite{Samarakoon_2017} as well as another highly frustrated variant, the Gamma model \cite{Samarakoon_2018}. 

\begin{figure}
\centering\includegraphics[width=0.75\textwidth]{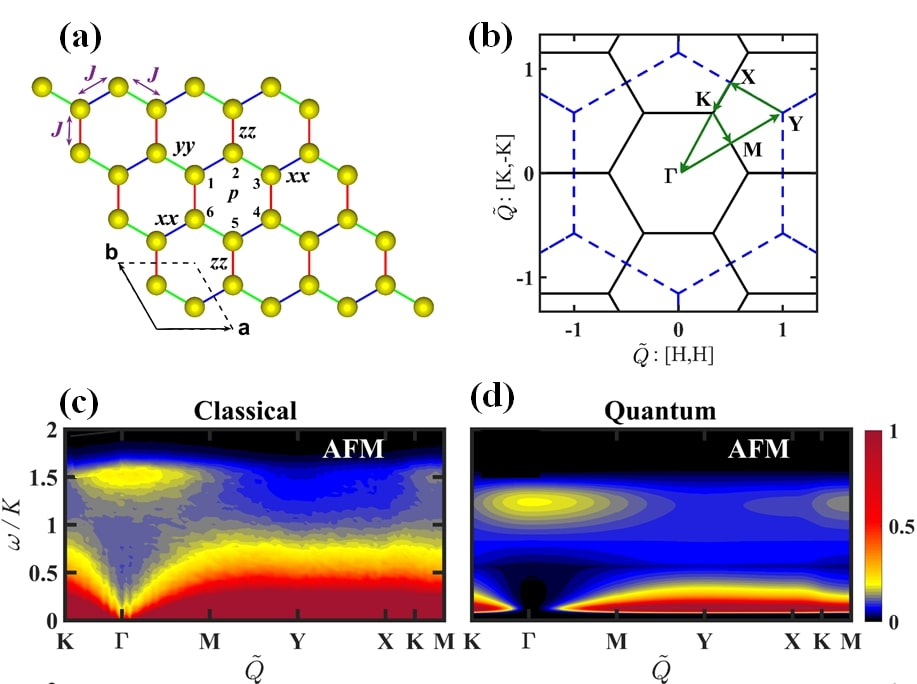}
	\caption{\textbf{Kitaev-Heisenberg Honeycomb Lattice:} (a) Schematic illustration of  the Kitaev-Heisenberg model consisting of the Heisenberg interaction $J$ and the compass-like nearest-neighbor Ising interactions, with the associated spin component for each bond depending on the bond orientations ($xx$, $yy$, or $zz$).
(b) The first BZ (solid line) and the second BZ (dashed line)  of the honeycomb lattice are shown.
The green arrows indicate a path connecting high-symmetry points in the reciprocal space (i.e., ${\bf K}-{\boldsymbol \Gamma}-{\bf M}-{\bf Y}-{\bf X}-{\bf K}-{\bf M}$), along which we evaluate $S(\mathbf{Q},\omega)$. $S\left(\mathrm{Q},\omega \right)$,
in (c) the classical limit $(S \to \infty)$ and  (d) quantum limit ($S=1/2$) of the pure Kitaev model ($J = 0$) at $T=0$. Panels (c) was obtained from LL simulations of the classical AFM Kitaev model. The figure is reproduced from \cite{Samarakoon_2018} and further details are given there.}
\label{flo:kitaev_sqw}
\end{figure}

The Kitaev-Heisenberg model with nearest-neighbor interactions is
\begin{equation}
    \mathcal{H}_{KH}=K \sum_{\nu=x,y,z}\sum_{\{i,j\}_{\nu}}S_{i}^{\nu}S_{j}^{\nu} + J \sum_{\{i,j\}} \textbf{S}_i \cdot \textbf{S}_j
\end{equation}
The index $\nu$ for the variables $i,j$ indicates the two neighboring sites connected by a $\nu\nu$ bond and $K$ and $J$ are the Kitaev and Heisenberg coupling strengths respectively (see Fig. \ref{flo:kitaev_sqw} (a)).
As shown in the Fig. \ref{flo:kitaev_sqw} (c) and (d), the differences and similarities in the dynamical structure factors of the spin-1/2 and the high spin ($S \rightarrow \infty$) classical Kitaev liquids were identified. Interestingly, the low-temperature and low-energy spectrum of the classical model exhibits a finite energy peak, which is the precursor of the one produced by the Majorana modes of the $S=1/2$ fully quantum mechanical model. 
The classical peak is spectrally narrowed compared to the quantum result and can be explained by magnon excitations within fluctuating one-dimensional manifolds (loops) which form in the ground state \cite{Samarakoon_2017}. Hence the difference from the classical limit to the quantum limit can be understood by the fractionalization of magnons propagating in one-dimensional manifolds. Moreover, the momentum space distribution of the low-energy spectral weight of the $S=1/2$ model follows the momentum space distribution of zero modes of the classical model ({\it c.f.} the animation of the ground state Supplementary Movie 1, ref. \cite{Samarakoon_2017} discussed at the end of \ref{computations}).

In the case of the Gamma ($\Gamma$) model the interactions are of a different form:
\begin{equation}
    \mathcal{H}_{\Gamma}=\Gamma \sum_{\alpha \neq \beta \neq \gamma}\sum_{\{i,j\}_{\nu}} \left( S_i^{\beta} S_j^{\alpha}
    + S_i^{\alpha}S_j^{\beta} \right).
\end{equation}
Again the semi-classical simulations \cite{Samarakoon_2018} capture the features of even the fully quantum correlations well and so provide a good basis for analyzing scattering before in depth study using full quantum calculations.

\section{Application of Machine Learning}\label{ml_application}

We now discuss the application of machine learning to neutron data and simulation. We do this with reference to the example of dysprosium titanate, Dy$_2$Ti$_2$O$_7$ (DTO). The model for this highly frustrated magnetic material is that of spin-ice (for a review of spin ice physics see \cite{spin_ice_review}) with exchange interactions up to 3rd nearest neighbors and long-ranged dipolar interaction on a pyrochlore lattice. Some of the simulations and data reduction are reported in \cite{Samarakoon:2020aa} however much of the discussion below is new to this paper.

\subsection{Data compression and latent spaces}

A typical neutron scattering data set from a single crystal is either 3D or 4D depending on whether it is a diffraction or spectroscopy experiment. Data consists of scattering events that are binned on regular grids in the three perpendicular reciprocal space axes ($\textbf{Q}_i$) or/and energy transfer ($\bf\omega$). The bin size is picked to be smaller than the instrument resolution and corrections to experimental factors such as detector efficiencies applied in the data reduction process. 

For DTO the data sets are 3D diffuse scattering (diffraction) measurements with of order $10^6$ bins. Neutron scattering experiments are generally data-limited so the machine learning is performed on the simulations instead. For this the Monte-Carlo simulator is run for the dipolar spin-ice Hamiltonian that includes exchange terms up to third-nearest neighbors:
\begin{equation}
\mathcal{H}=\sum_{\alpha=1,2,3,3'} J_{\alpha} \sum_{{\lbrace i,j \rbrace}_{\alpha}} 
S_i \cdot S_j + 
\mathcal{D} r_1^3 \sum_{\lbrace i,j \rbrace} 
\left\lbrack \frac{S_i.S_j}{| r_{ij}|^3}
-\frac{3({\bf S}_i.{\bf r}_{ij}  ).({\bf S}_j.{\bf r}_{ij}  )}{| r_{ij} |^5} \right\rbrack 
\end{equation}
where ${\bf S}_i$ is the Ising spin of the $i^{th}$ ion. 
The model includes first, second, and two different third nearest neighbors with interaction strengths, $J_1$, $J_2$, $J_3$ and $J_{3'}$ respectively. 
There is also a dipolar interaction with strength $\mathcal{D}$, which couples the $i^{th}$ and the $j^{th}$ spins according to their displacement vector $r_{ij}$. In this section, we are exploring over $J_2-J_3-J_{3'}$ space at fixed values of  $\mathcal{D} = 1.3224$~K and $J_1=3.41$~K, which are determined to high accuracy from prior work. 

The initial training data (used in the optimization below) uses 8100 3D volumes of neutron scattering structure factors, $\mathcal{S}^{sim}({\bf Q})$, on evenly spaced grids of $30 \times 30 \times 9$ over the  $J_2-J_3-J_{3'}$ space. The system size used comprises 1024 spins and simulations are performed at the same temperature as the experiment (680 mK). The collected thermalized spin configurations  were Fourier transformed into reciprocal space to generate a simulated structure factor, $\mathcal{S}^{sim} ({\bf Q})$. Other experimental factors such as magnetic form factor of Dy$^{3+}$ and the neutron scattering polarization factor are accounted for so model and experiment can be compared.

\begin{figure*}
\centering\includegraphics[width=0.99\textwidth]{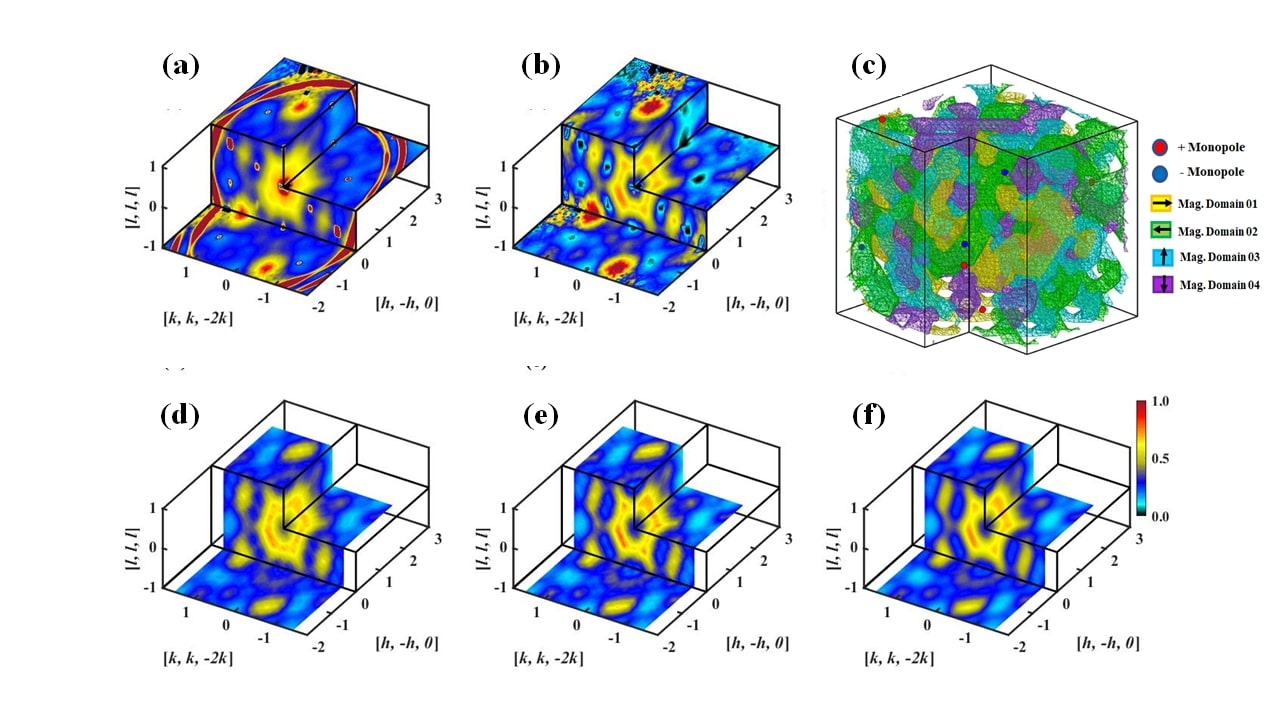}
	\caption{\textbf{Machine learning applied to Dy$_2$Ti$_2$O$_7$:} Machine learning using trained autoencoders results in visualization of glass formation in a cooled spin liquid. High performance computing is used to generate a large number of simulations of neutron experiments in a high dimensional model space $\{ \mathcal{H}(J)\}$. Optimization of the model parameters to the measured data is undertaken in the autoencoder latent space which is robust to background and instrumental artifacts.  Panel (a) is a rendering of untreated 3D diffuse scattering measurement of the spin correlations $\mathcal{S}^{exp}({\bf Q})$ in the gauge spin liquid Dy$_2$Ti$_2$O$_7$ measured as one of 40 data sets on CORELLI over a three-day experiment covering a range of temperatures and fields. Rings of scattering from the sample mounting are visible. Panel (b) shows the data after conventional manual treatment to remove non-magnetic contributions. (c) Visualization of the model in real space showing monopole localization and intertwined order formation that constitutes the emergent low temperature glass phase. (d) The untreated data (panel (a)) processed directly by the NLAE. This results in denoising and removal of experimental artifacts. The result is almost indistinguishable from (e) the same processing performed on treated data (from panel (b)). The result is comparable to full optimization where (f) shows the best model which provides optimized model parameters from the 3D data set. }
\label{flo:QFIa}
\end{figure*}

Each $\mathcal{S}^{sim}({\bf Q})$ is matched to the binning of the data, see Fig. \ref{flo:QFIa}. For the purposes of compression, each pixel is a single dimension. For automatic feature extraction we compare dimensionality reduction using PCA and NLAEs. 

\subsubsection{Principle component analysis}

Principal component analysis is a commonly used linear dimensionality reduction technique. It involves computing the principal components of the input distribution and linearly transforming from the original basis \cite{PCA_Jolliffe}. The basis is equivalent to the number of ${\bf Q}$-points (mid-point of bins) in this case. A single value decomposition (SVD) is used to calculate principal components (latent variables which constitute a latent space) of the data structure comprising a set of $\{ \mathcal{S}^{sim} ({\bf Q}) \}$ evaluated over $\{ {J} \}$,

\begin{figure}
\centering\includegraphics[width=1\textwidth]{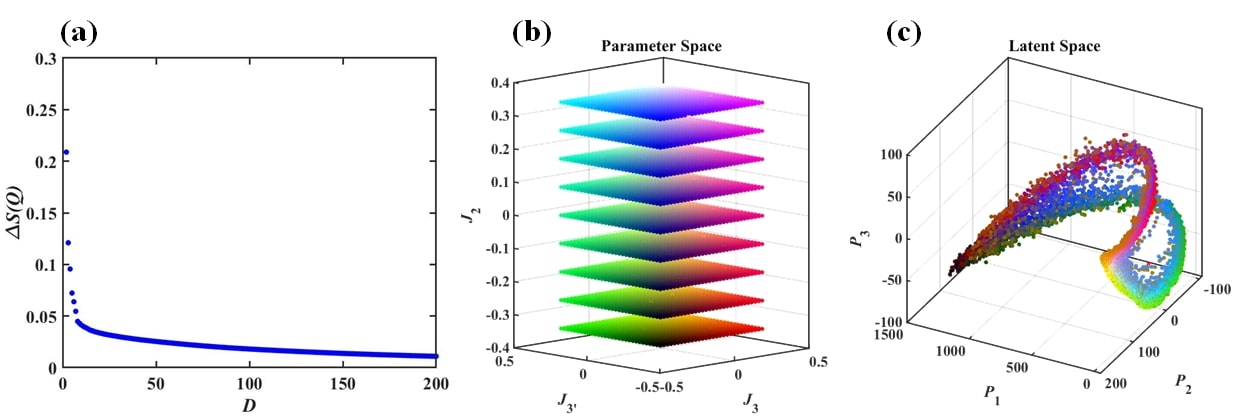}
	\caption{\textbf{Principal component analysis:} (a) Total reconstruction error, $\Delta \mathcal{S}({\bf Q})$ with PCA dimension, $D$. (b) Three-dimensional parameter space of $J_2-J_3-J_{3'}$ sampled in a coarse grid and colored by treating its coordinate as an RGB value. (c) Mapping parameter space, $\{ J \}$ to 3D latent space, $\{ P \}$. The colors are consistent in both (b) and (c); so the transfer function ${P}=L^{PCA} ({J})$ can be easily seen. In this latent space representation, a broad phase in parameter space appears as a dense point in latent space and a higher order phase transition / crossover appears as lines of points connecting dense clusters. First-order phase transitions appear as isolated clusters and discontinuities in the latent space projection. 
	}
\label{flo:pca_figure}
\end{figure}

\begin{figure}
\centering\includegraphics[width=1\textwidth]{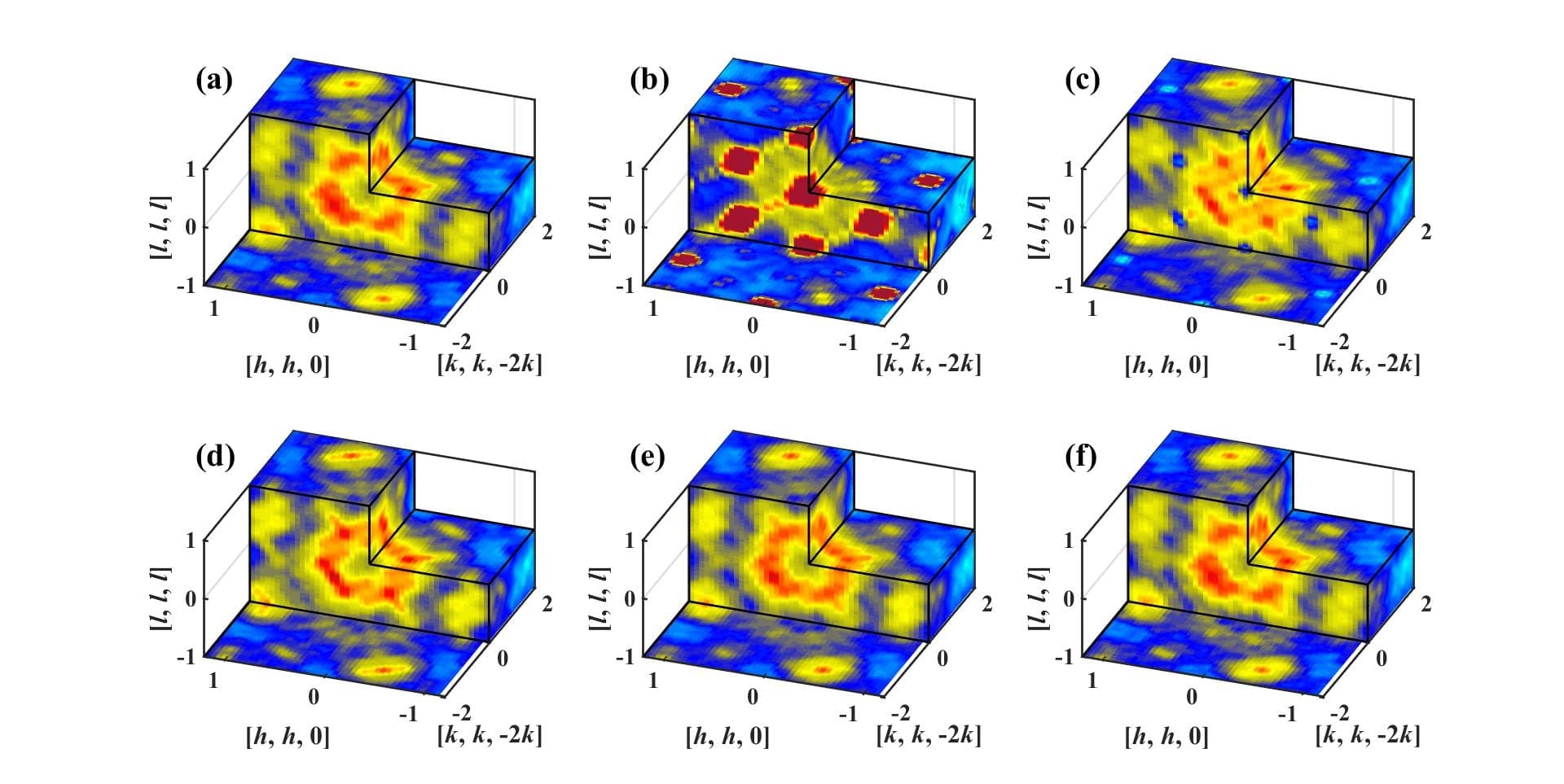}
	\caption{\textbf{Reconstruction from principal components:} The fidelity to which input data is recovered with number of principal components is shown. Panel (a) shows a randomly selected simulation from the training data set to check the reconstruction ability of the PCA. Reconstructed structure factors, $\mathcal{S}_n^{pca} ({\bf Q})$ are shown in panels (b-f) generated by keeping only (b) one ($n=1$), (c) two ($n=2$), (d) four ($n=4$), (e) $n=20$ and (f) $n=100$ components. 
    
	}
\label{flo:pca_reconstruct}
\end{figure}

To show the effectiveness of PCA for compressing the data, we take 2000 of the training data sets and perform analysis on them. The first few dimensions are most significant with only three independent dimensions having a percentage variance higher than 5\% and adequate enough to distinguish main features of $ \mathcal{S}^{sim}({\bf Q}) $.
The fidelity of reproduction of the simulations versus number of components, $D$ is shown in Figure \ref{flo:pca_figure}(a) by the total reproduction error:
\begin{equation}
\Delta \mathcal{S}({\bf Q}) = \sum_{k}\sum_{{\bf Q}} (\mathcal{S}^{sim}_{k}({\bf Q}) - \mathcal{S}^{pca}_{k}({\bf Q }))^2
\label{eq:TotalError}
\end{equation}\label{eq:error}
where  $\mathcal{S}^{pca}_{k}({\bf Q})$  is the reconstructed structure factor of the $k^{th}$ training data by considering $D$ number of PCA dimensions with highest variance. As the number of components is increased the reproduction gets significantly better and by $n=4$ reproduction is seen to significantly improve the comparison with the simulation. However, by $n=20$ there is still some discernible improvement (see Fig. \ref{flo:pca_reconstruct}).

Before moving onto the nonlinear autoencoder, we consider how to visualize the phases and transformations that happen in the interaction space. Fig. \ref{flo:pca_figure} panels (b) and (c) show the mapping from the interaction space of the model into the first three principal components which form a simplified latent space. Fig. \ref{flo:pca_figure} (c) shows a locus of interaction coordinates transformed into the latent space. These group into high density points which are indicative of distinct physical phases. This shows the physical information content of the neutron scattering data provides an excellent discriminator for distinguishing phases in materials and accounts for its power as an experimental technique. Between these dense locations there are connecting points which show kinks and inflections. These indicate the presence of phase transitions and the nature of the transition (first and second order or crossovers).

\subsubsection{Non-linear autoencoders}

\begin{figure}
\centering\includegraphics[width=.8\textwidth]{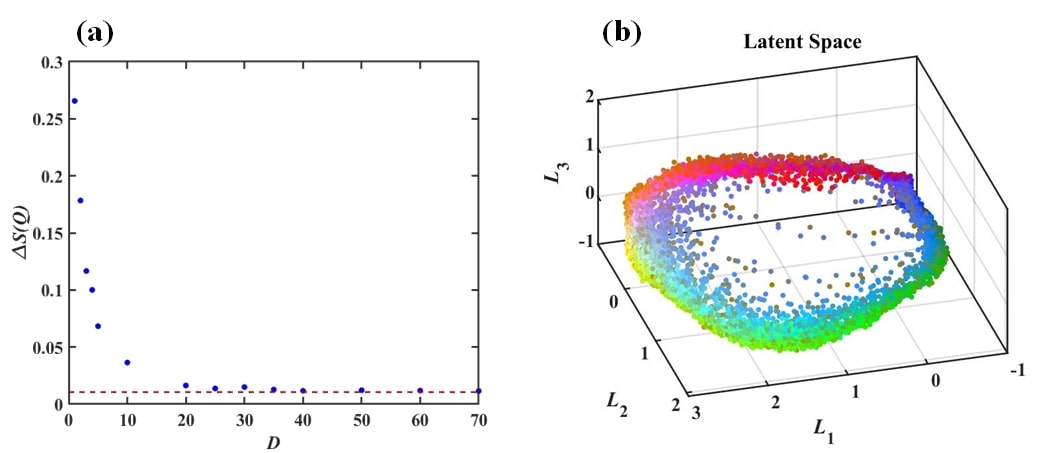}
	\caption{\textbf{Reconstruction from principal components:} 
	(a) Total reconstruction error, $\Delta \mathcal{S}({\bf Q})$ versus Non-Linear Autoencoder (NLAE) dimension, $D$. (b) Mapping parameters space, $\{ {J} \}$ to 3D latent space $\{ L \}$. The colors follow Fig. \ref{flo:pca_figure} (b); the transfer function, $\{ L \} =\mathcal{L}^{NLAE} (\{ {J} \})$ can be easily seen. The parameter space manifold from the non-linear autoencoder is more stretched than in the case of PCA \ref{flo:pca_figure}(c) by creating more distinction between points through the nonlinearity.
	}
\label{flo:nlae_analysis}
\end{figure}

Now we consider the use of non-linear autoencoders for dimensionality reduction.  Autotoencoders are neural networks trained to compress then decompress images through a latent layer (dimension $D$) to reproduce the original image. While linear autoencoders essentially undertake principal component analysis, further compression is possible with NLAEs. We undertake the same analysis as above using a NLAE with a logistic (sigmoid) activation function, see \cite{Samarakoon:2020aa}. The NLAE encodes the simulations into the compressed latent space representation $\mathcal{L}_D=(L_1,L_2,...,L_D)$  and then decodes it to an output $\mathcal{S}^{AE}\left(\textbf{Q}\right)$ capturing the essence of the input $\mathcal{S}\left(\textbf{Q}\right)$ while removing irrelevant noise and artifacts.  The hyper-parameter $D$ can be optimized by considering the total reproduction error $\Delta \mathcal{S}({\bf Q})$, see Fig. \ref{flo:nlae_analysis} (a), which is defined as in eq.\ref{eq:TotalError} but replacing $\mathcal{S}^{pca}_{k}({\bf Q})$ with $\mathcal{S}^{AE}_{k}({\bf Q})$. $\Delta \mathcal{S}({\bf Q})$ converges by around $D=20$ for the non-linear autoencoder whereas the PCA has not fully converged even at $D=100$ demonstrating the superior performance of the NLAE at dimensionality reduction. 

It is advantageous to reduce the dimensionality of the latent space as far as possible. A smaller dimensionality $D$ aids visualization and extraction of meaningful information from the data structures. It also helps in building better generative models to predict $\mathcal{S}({\bf Q})$ for given parameters; capabilities important for application (discussed later in subsection \ref{generative}). The NLAE used here is a simple Neural Network architecture; a single hidden layer, fully-connected network. Further reduction of dimensionality may be achievable by implementing multi- or variational-layer architectures. This is out of the scope of the present paper but will be explored in a future publication.


A NLAE was trained with $D=20$, the determined optimum dimensionality: Fig. \ref{flo:nlae_analysis} (b) shows the manifold of interaction coordinates transformed into the first three NLAE latent space dimensions with highest variance. For the PCA, Fig. \ref{flo:pca_figure}(c), the interaction manifold is a highly folded 2D plane while for the NLAE it is unfolded and stretched. The NLAE has increasing the pair-wise distance between points in the latent space and provides a better error measure for optimization and more effective categorization of phases. 

\subsection{Phase identification and mapping}

Identifying regions of distinct physical behavior and the automatic mapping of phase diagrams in materials are essential problems facing materials science. The competition between phases and stabilizing desired states is important to both application and understanding, It is natural to use heirarchical clustering on the simulated data to undertake phase mapping. The scattering cross-section $\mathcal{S}^{sim} ({\bf Q})$ has been shown to be information rich with features indicative of underlying phases so can be expected to be well suited to discriminating and classifying states (see Fig. \ref{flo:fitting} (a)). 

\begin{figure}
\centering\includegraphics[width=1\textwidth]{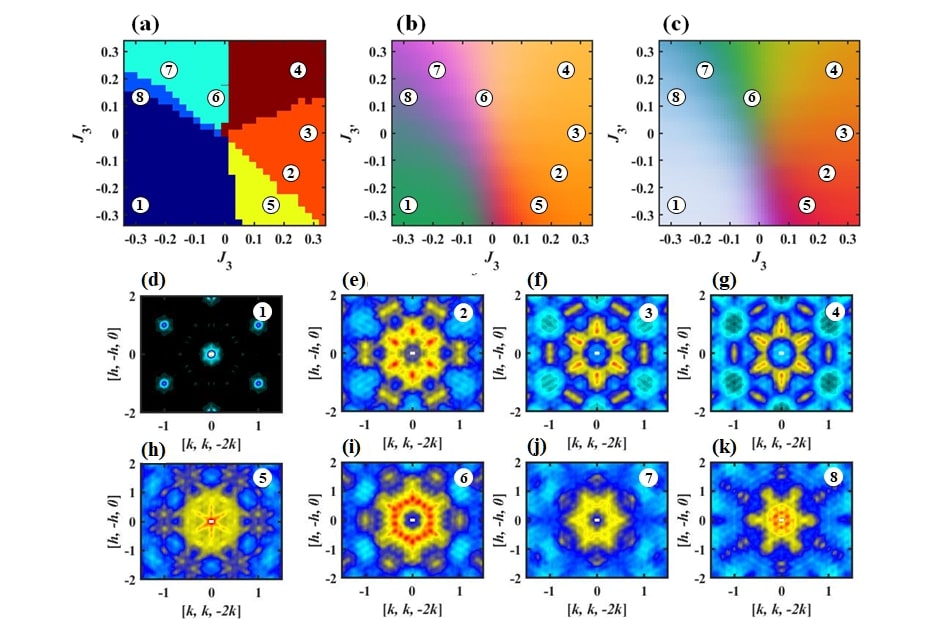}
	\caption{\textbf{Mapping of phases from simulations:} 
	 Comparison of phase maps generated through different approaches. Here, we have considered three techniques to create phase maps using information extracted from collected $\mathcal{S}^{sim} ({\bf Q})$. (a) Applying hierarchical clustering directly on latent space variables from the NLAE. (b) Using the three PCA dimensions with highest variance.  (c) Using the three NLAE dimensions with highest variance. For (b) and (c), the latent space coordinates have been converted to an RGB color code. For panel (a), the number of clusters was pre-defined in an automated fashion. Panels (d)-(k) show slices of $\mathcal{S}^{sim} ({\bf Q})$ at corresponding positions in (a)-(c) indicated by their numerical code.
	}
\label{flo:phase_diagram}
\end{figure}

First we classify and map out the domains of different features with $\{ {J} \}$ using agglomerative hierarchical clustering \cite{Zhang_2012}. 
The clustering requires a pairwise metric and a squared distance between $\mathcal{S}^{sim} ({\bf Q}, \{J\})$ was applied to the training data sets. Figure \ref{flo:fitting} (a) shows that this metric makes visible different phases of behavior. However, as the $\mathcal{S}^{sim} ({\bf Q},\{J\})$ have already been compressed, the latent space from either the PCA or NLAE can be used and is computationally much faster while also reducing effects of statistical noise from the simulations. A phase diagram generated using this approach with an autoencoder latent space \cite{Samarakoon:2020aa} is shown in Fig. \ref{flo:phase_diagram}(a). The different colored regions indicate clusters and are broadly indicative of distinctive physical behavior. Not all the phases are truly distinct with thermodynamic phase transitions between and some are in fact crossovers in behavior. Representative correlation functions are indicated in Fig. \ref{flo:phase_diagram} (d-k). 

A second approach, which is less rigorous, is simply to display the latent space in ways that make potential phases and transitions more apparent. Figs. \ref{flo:phase_diagram}(b) and (c) show the three most significant components of the PCA and NLAE latent spaces, respectively. In these plots phases can already be discerned and the NLAE in particular shows a close relation with the heirarchical clustering.

There is considerable scope for improvement in mapping phases. First, undertaking adaptive searches that delineate phase boundaries are clearly useful. Second, using generative models to speed up mapping in high dimensional spaces will be needed. Third, undertaking analysis directly on the spin configurations and combining with other observables such as other susceptibilities should be effective. 

\subsection{Parameter optimization from data}\label{optimization}

So far we have shown how large scale neutron scattering simulations in conjunction with machine learning methods can compress, cluster, and classify information leading to new insights. The next important goal is how data analysis and extraction of the model and its parameters can be achieved.

\begin{figure*}
\centering\includegraphics[width=0.8\textwidth]{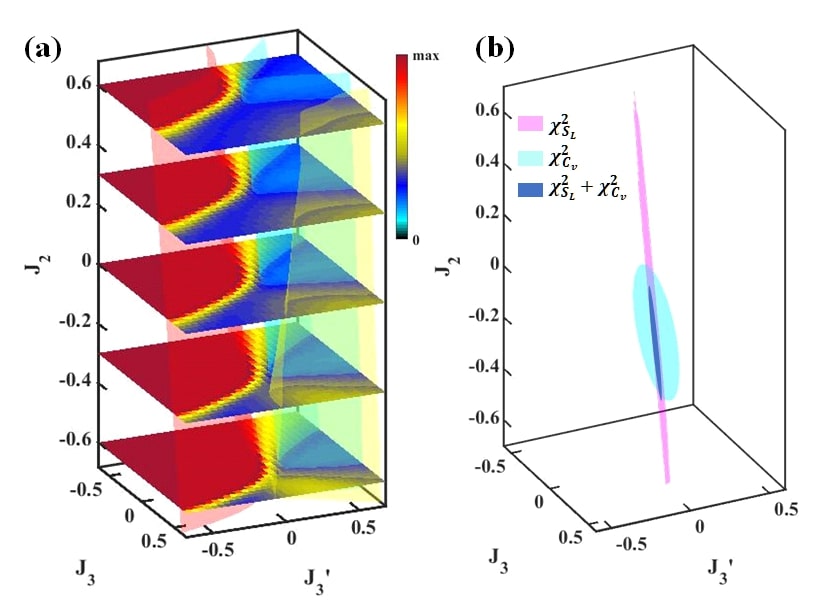}
	\caption{\textbf{Cost functions for optimization:} (a) Direct distance squared between $\mathcal{S}^{exp}({\bf Q})$ and $\mathcal{S}^{sim}({\bf Q})$, $\chi^2_{\mathcal{S}({\bf Q})}$ plotted as a volume. The colors correspond to the value of $\chi^2_{S(Q)}$ while the surfaces are at fixed values of $\chi^2_{\mathcal{S}({\bf Q})}$. (b) The uncertainties of (magenta) $\chi_{S_L}^2$, defined as the distance between the latent space representation of the $S(Q)$ data, (cyan) $\chi_{C_v}^2$, defined as the distance between measured and simulated heat capacities and (blue) the multi-objective error measure $\chi_{multi}^2$ = $\chi_{S_L}^2$+$\chi_{C_v}^2$. The error measure, $\chi_{S_L}^2$, has been shown to perform better than the regular $\chi^2_{\mathcal{S}({\bf Q})}$ \cite{Samarakoon:2020aa}.
	}
\label{flo:fitting}
\end{figure*}

Conventional fitting procedures involve directly calculating the scattering from the model and undertaking a chi-squared comparison between model ($\mathcal{S}^{sim}({\bf Q})$) and 
measurement ($\mathcal{S}^{exp}({\bf Q})$), denoted by $\chi^2_{\mathcal{S}({\bf Q})}$. An onerous part of this process is the data treatment for the unavoidable contamination by experimental artifacts. This needs to be masked out and corrections to account for non-magnetic backgrounds applied. In the case of the very large data flows coming from contemporary instrumentation, manual approaches to treatment and 
even exploring the 3D and 4D volumes of each data set is impractical. Even if careful data treatment and modeling were possible, conventional methods still have difficulty finding optimal solutions due to the biasing from remaining artifacts and statistical noise in data and simulations \cite{Samarakoon:2020aa}. Recent developments suggest that machine learning based approach can overcome these problems.  

To apply unsupervised machine learning to the analysis of DTO neutron data, the NLAE from above was used. 
For the analysis and optimization of the model ${\mathcal{H}\{{J}\}}$ random sampling of the Hamiltonians was undertaken to iteratively build up an accurate representation around the region of interest. An error measure defined as the least squares difference in the {\it latent space}, $\chi^2_{S_L}$  is found to be more sensitive to features in the data and less biased by artifacts \cite{Samarakoon:2020aa} than the conventional $\chi^2_{\mathcal{S}({\bf Q})}$ between data and model. Indeed this autoencoder-based error measure is more robust to stochastic noise, i.e., allows more precise estimation of $\{J\}$. A comparison of the two error measures, $\chi^2_{\mathcal{S}({\bf Q})}$ and $\chi^2_{S_L}$  is give in Fig. 4 (d) of ref. \cite{Samarakoon:2020aa}. A low-cost estimator of $\chi^2_{S_L}$, ${\hat\chi}^2_{S_L}$ was found using Gaussian Process regression. The optimization involves randomly selected Hamiltonians for inclusion in the dataset, subject to the constraint,  that ${\hat\chi}^2_{S_L}$ is below a cut-off parameter $c_L$.. The cut-off decreases monotonically so that later iterations in the optimization procedure are focused on regions where the latent space chi-squared is smallest.  
After sufficient iterations the best optimization is found.



Fig. \ref{flo:QFIa} shows the treatment of a typical dataset from DTO by the NLAE. Raw data without any correction of the experimental background are shown in Fig. \ref{flo:QFIa} (a). Bragg peaks and scattering in the form of powder rings from the crysostat (sample environment) are evident. Fig. \ref{flo:QFIa} (b), meanwhile, shows the data after being manually processed to remove these artifacts. It is not possible to correct completely for such scattering and that plus difficulties such as detector efficiencies and multiple scattering mean some artifacts remain. Fig. \ref{flo:QFIa} (d) and (e) are the NLAE filtered data for the untreated and treated experimental data respectively.  The NLAE has successfully filtered out the artifacts even for the untreated data set removing the need for lengthy manual pre-processing. The $\mathcal{S}^{sim}({\bf Q})$  for the optimized parameters is also shown in Fig.\ref{flo:QFIa} (f) in comparison to the raw and filtered data. This then allows automation of fitting and rapid analysis of experiments in cases where training of NLAEs can be performed beforehand. 


Using a mapping from the latent space to the model means the NLAE performs inversion of the scattering to the model. In other words, it solves the inverse scattering problem when the training data is sufficiently realistic and comprehensive enough in scope. Further, in the case of DTO the real space configuration that the scattering corresponds to can be recovered from the simulations allowing the state of the material to be visualized, see Fig. \ref{flo:QFIa}(c).   

A final consideration is the combination of multiple techniques and impact on broader experimental strategy and planning. The optimization results in regions of uncertainty of the model parameters. These can be narrowed down further by changing experimental variables such as temperature or applied magnetic field and remeasuring. Also, data such as heat capacity and susceptibility can be used to localize the best parameters (see the Fig. \ref{flo:fitting}(b)). Modeling and use of NLAEs together can help to plan strategies for most effectively undertaking measurements as well as suggesting modifications of materials to stabilize desired phases.  

\subsection{Generative Models}\label{generative}
A generative model (GM) can effectively replace simulations for calculating physical properties such as neutron scattering when correctly trained. 
Building accurate generative models allows us to comprehensively study high-dimensional parameter spaces and search for exotic phases without paying the high-computational cost of very large numbers of direct simulations. In addition the GM can be made 
available for future use and to other researchers. 
There is uncertainty in prediction of the GM but these can be verified for notable results with the simulations. 
Moreover, GMs can be improved over time as more simulation data is collected by retraining. 

In subsection \ref{optimization}, low cost estimators for  $\chi^2_{S_L}$ were mentioned. Similarly, a ML model ($NN_L$) between the model parameters $\{J\}$ and latent space parameters $\{L\}$ can be constructed and by linking this network with the decoder network of the NLAE a simple GM architecture can be implemented. We have successfully done this for magnetic systems that predict the static neutron structure factors and used the GM to locate and obtain the neighboring phases in the high-dimensional parameters spaces. Technical details are out of the scope of this paper and will be the topic of an upcoming publication. While the concept of generative models is similar to N-dimensional-interpolation (where $N$ is the basis of $Q$ points), the dimensionality in this case is too large to compute effectively. Thus, using compression is crucial for building these models.

\section*{Discussion}
As well as application to a wide range of disordered materials and interpretation of diffuse scattering, the above approaches should be extendable to other classes of scattering problems. We have explored single crystal inelastic neutron scattering and the approach here provides in principle a powerful new approach to this. Inelastic neutron scattering from pulsed neutron sources collect 4D data that is proving very hard for even experienced scientists to interpret and analyze. Training neutral networks to interpret and process the data gives a potentially highly automatable new tool for undertaking the analysis. 

Producing single crystals suitable for study is time intensive and is often not even possible. Scattering from powders avoids these difficulties but the orientational disorder averages out the scattering resulting in loss of information. The powder spectra are very hard to directly interpret and conventional fitting techniques face serious problems with identifying the extent that any model extracted is unique. Again, machine learning approaches combined with high performance computing should provide a way of addressing this and therefore open up the use of powders. This could dramatically speed up identification and understanding of materials.    

Mapping out magnetic phase diagrams in an automated way is of great importance. In many materials now under consideration the large number of potential terms in the model makes this a formidable problem. The approach above using auto-encoders can also be extended to be optimized for phase detection and identification of boundaries and speeded up using generative models. Locating materials in a higher dimensional landscape of phases aids materials design and prediction by identifying strategies to modify materials to stabilize desired phases.

As we have seen here, $\mathcal{S}({\bf Q})$ and $\mathcal{S}({\bf Q},\omega)$ are particularly information rich physical observables so are excellent discriminators for classes of physical behaviors and phases. As such the auto classification is of great importance to the physical understanding of the system and can be used for both ordered and disordered materials including topological materials. Because of the large dimensional space involved in $\{ \mathcal{H}( \{J\}) \}$ machine learning approaches are required and this represents a new capability for researchers. It can be foreseen that electronic structure calculations should be able to give indications of the most important potential interactions in the material and rough estimates of these and so can be integrated into the training process.

To understand more broadly the applicability of semi-classical methodologies we consider principal ways quantum effects manifest in the spectra. Higher order perturbative effects in spin wave theory cause renormalizations in the energy scale of the dispersions. These can be absorbed into the exchange couplings and correction factors assessed by comparison to full quantum mechanical computation. A further effect is quantum gap formation characteristic of quantum spin liquid formation or dimerization. 
Quantum gaps are small compared to the energy scale of the dispersion. As the gaps are due to entanglement associated with singlet formation the semi-classical approaches are unable to capture these features and ML approaches trained on spin-wave or LL based simulations can only hope to locate roughly the underlying model with further in-depth quantum simulation being required to capture this aspect. For temperatures above the energy scale of the gaps and interactions, the correlations should be more similar to the semi-classical simulations. This is seen in the Heisenberg and honeycomb examples above, subsection \ref{examples} and measurements at elevated temperatures can be undertaken. 

Continua are another manifestation of quantum fluctuations. Both multi-spin-wave excitations and fractionalization of quantum numbers in magnets produce low-temperature continua without classical counterpart that will not be recognized by LL or linear spin-wave based training. However, the general distribution of scattering weight and higher temperature behavior will be matched so again a localization of the model can be be expected.

In magnetic materials involving itinerant electrons, multiple crystal field levels, or dimerized singlets multiple quantum levels and degrees of freedom are involved that require other approaches beyond those here. However, approaches to these suitable for adapting to machine learning are realistic and should be pursued.  

Overall, we think that machine learning could have profound and far reaching impact on neutron scattering from magnetic materials. This will extend to accelerating the modeling of systems including quantum effects too. Much of our considerations apply equally well to soft matter, solid state chemistry, and biological systems where large scale modeling is also available. Some prelimary use of machine learning is happening with neutrons in these areas too \cite{ml_review}. Machine learning naturally extends into autonomous steering of experiments and the use of multi-modal data and needs to be considered in the context of optimal use of expensive and complex experiments and design of new instrumentation. 

\section*{Conclusion}

The use of machine learning to neutron scattering and magnetism are at a very early stage. Many promising results however are coming out which show that such approaches can make very significant impact on experiment and analysis. The work shown here should be regarded as only a start to the use of machine learning and there are enormous potential developments in simulations and for more sophisticated algorithms on the horizon. Machine learning combined with high performance simulations represent a new capability that we believe should be vigorously pursued.

\section*{Summary}
The application of machine learning to neutron scattering was demonstrated for magnetic materials. Simulation approaches using Landau Lifshitz and Monte Carlo methods were considered and shown to provide effective modeling of magnetic structures and dynamics in a wide range of materials. Using simulations of neutron scattering experiments as training data for machine learning showed that a high degree of compression was successfully achieved and that non-linear autoencoders were very effective. Different approaches to extracting phases and transitions were demonstrated and the use of optimization of parameters using latent spaces was found to be advantageous over conventional fitting approaches. Overall machine learning was demonstrated to be able to automate complex data analysis tasks including the inversion of neutron scattering data into a model.

\section*{Acknowledgements}

The authors benefited enormously from highly stimulating discussions and collaborations with Cristian Batista, Kipton Barros and Vickram Sharma. The computer modeling presented in this paper used resources of the Oak Ridge Leadership Computing Facility, which is supported by the Office of Science of the U.S. Department of Energy under contract no. DE-AC05-00OR22725. AS was supported by the DOE Office of Science, Basic Energy Sciences, Scientific User Facilities Division. The work by DAT is supported by the Quantum Science Center (QSC), a National Quantum Information Science Research Center of the U.S. Department of Energy (DOE).

\section*{References}
\bibliographystyle{vancouver}
\bibliography{mlphasediag}

\end{document}